\documentclass{mem}
\usepackage{natbib}
\usepackage{txfonts}
\usepackage{balance}
\usepackage{graphicx}
\usepackage[a4paper,breaklinks,dvipdfm]{hyperref}
\idline{00}{1}
\begin{document}

\title{The relation between bar formation, galaxy luminosity, 
  and environment}

\subtitle{}

\author{
E. M.\, Corsini\inst{1,2} \and
J.\, M\'endez-Abreu\inst{3,4} \and
R.\, S\'anchez-Janssen\inst{5} \and 
J. A. L.\, Aguerri\inst{3,4} \and \\
S.\, Zarattini\inst{1,3,4}}

\offprints{E. M.\, Corsini}
 
\institute{
Dipartimento di Fisica e Astronomia `G. Galilei', 
Universit\`a di Padova, 
Padova, Italy
\and
INAF--Osservatorio Astronomico di Padova, 
Padova, Italy
\and
Instituto de Astrof\'isica de Canarias, 
La Laguna, 
Spain
\and
Departamento de Astrof\'isica, 
Universidad de La Laguna,  
La Laguna, 
Spain
\and 
European Southern Observatory, 
Santiago, Chile\\
\email{enricomaria.corsini@unipd.it}}

\authorrunning{Corsini et al.}

\titlerunning{Bar formation, galaxy luminosity, and environment}

\abstract{
We derive the bar fraction in three different environments ranging
from the field to Virgo and Coma clusters, covering an unprecedentedly
large range of galaxy luminosities (or, equivalently, stellar
masses). We confirm that the fraction of barred galaxies strongly
depends on galaxy luminosity. We also show that the difference between
the bar fraction distributions as a function of galaxy luminosity (and
mass) in the field and Coma cluster are statistically significant,
with Virgo being an intermediate case.
We interpret this result as a variation of the effect of environment
on bar formation depending on galaxy luminosity. We speculate that
brighter disk galaxies are stable enough against interactions to keep
their cold structure, thus, the interactions are able to trigger bar
formation. For fainter galaxies the interactions become strong enough
to heat up the disks inhibiting bar formation and even destroying the
disks. Finally, we point out that the controversy regarding whether
the bar fraction depends on environment could be resolved by taking
into account the different luminosity ranges of the galaxy samples 
studied so far.
}

\maketitle{}

\section{Introduction}
\label {sec:introduction}

Galaxy bars spontaneously form due to instabilities in dynamically
cold disks.  The growth rate of bars depends on the halo-to-disk mass
ratio and the velocity dispersions of the disk and halo and bars grow
faster in massive and cold disks \citep[see][for a
  review]{Athanassoula2012}.
Moreover, environmental processes can regulate the life cycle of bars
contributing both to their development by forcing disk instabilities
and to their destruction via disk heating.
A variety of methods to detect
bars in galaxy disks and measure the local galaxy density have been
adopted, but the results on the correlation between the bar fraction
and environment are controversial \citep[see][and references
  therein]{MendezAbreu2012}.

Here, we derive the bar fraction in three different galaxy
environments ranging from the field to the Virgo and Coma clusters.
The unprecedentedly large range of luminosities (or, equivalently,
stellar masses) covered by the different galaxy samples we investigate
allows us to distinguish the effects of environment in heating galaxy
disks from those triggering bar formation.

\section{Galaxy sample, identification of disks, and detection of bars}
\label{sec:sample}

Four galaxy samples were selected in order to analyze three different
galaxy environments (i.e., the field, Virgo cluster, and Coma
cluster):

\begin{itemize}

\item field1 sample: 2389 field galaxies with magnitude
  $-24\,\lesssim\,M_r\,\lesssim\,-20$ from \citet{Aguerri2009};

\item field2 sample: 352 field galaxies with
  $-21\,\lesssim\,M_r\,\lesssim\,-13$ from \citet{SanchezJanssen2010};

\item Virgo sample: 588 Virgo galaxies with
  $-22\,\lesssim\,M_r\,\lesssim\,-13$ from Zarattini et al. (in
  prep.);

\item Coma sample: 169 Coma galaxies with
  $-23\,\lesssim\,M_r\,\lesssim\,-14$ and located mainly in the
  cluster center from \citet{MendezAbreu2010}.

\end{itemize}

The $g$ and $r$ apparent magnitudes and axial ratios $b/a$, of all the
galaxies were retrieved from Sloan Digital Sky Survey (SDSS) III
(\citealt{Aihara2011}). We considered only the galaxies with
$b/a\,>\,0.5$ in order to deal with projection effects.
The galaxy images of field2, Virgo, and Coma samples have a similar
spatial resolution ($r_{\rm bar}\simeq150$ pc). Indeed, the farthest
galaxies (i.e., those in the Coma sample) were analyzed by studying
their {\em Hubble Space Telescope\/} images.  The mean resolution of
the field1 sample is $r_{\rm bar}\,\simeq\,1.3$ kpc. However, since
the field1 galaxies are the largest galaxies in our sample, bars
smaller than the resolution limit should be considered as nuclear
bars.
We computed the stellar mass from the $g-r$ color following
\citet{Zibetti2009} to avoid a possible color bias due to cluster
galaxies being on average redder than in the field.

We adopted the morphological classification of the galaxies in the
SSDS-Data Release 7 spectroscopic sample given by
\citet{HuertasCompany2011}. Due to the incompleteness of the SDSS
spectroscopic sample and since several of our sample galaxies do not
have any spectroscopic information, we remained with 1604, 336, 228,
and 44 disk galaxies in the field1, field2, Virgo, and Coma sample,
respectively.
Following \citet{Aguerri2009} and \citet{MendezAbreu2010}, we detected
the presence of bars by visually inspecting the galaxy images. We
classified all the galaxies into strong barred, weakly barred, and
unbarred.

\section{Results}
\label{sec:results}

For each galaxy sample we derived the {\em ordinary\/} bar fraction
$f_{\rm D}$ (calculated only for the disk galaxies) and the {\em
  overall\/} bar fraction $f_{\rm T}$ (calculated for all the galaxies
independently of their Hubble type).
Since bars can only be triggered in disks, $f_{\rm T}$ combines the
luminosity distribution of disk galaxies with their probability of
having a bar overcoming the problem of the identification of disk
galaxies.  This is always a major concern in morphological
classifications dealing with the measurement of bar fraction.  $f_{\rm
  T}$ allows us to probe a larger range of luminosities than $f_{\rm
  D}$.

\begin{figure*}[!ht]
 \centering
 \includegraphics[width=0.45\textwidth]{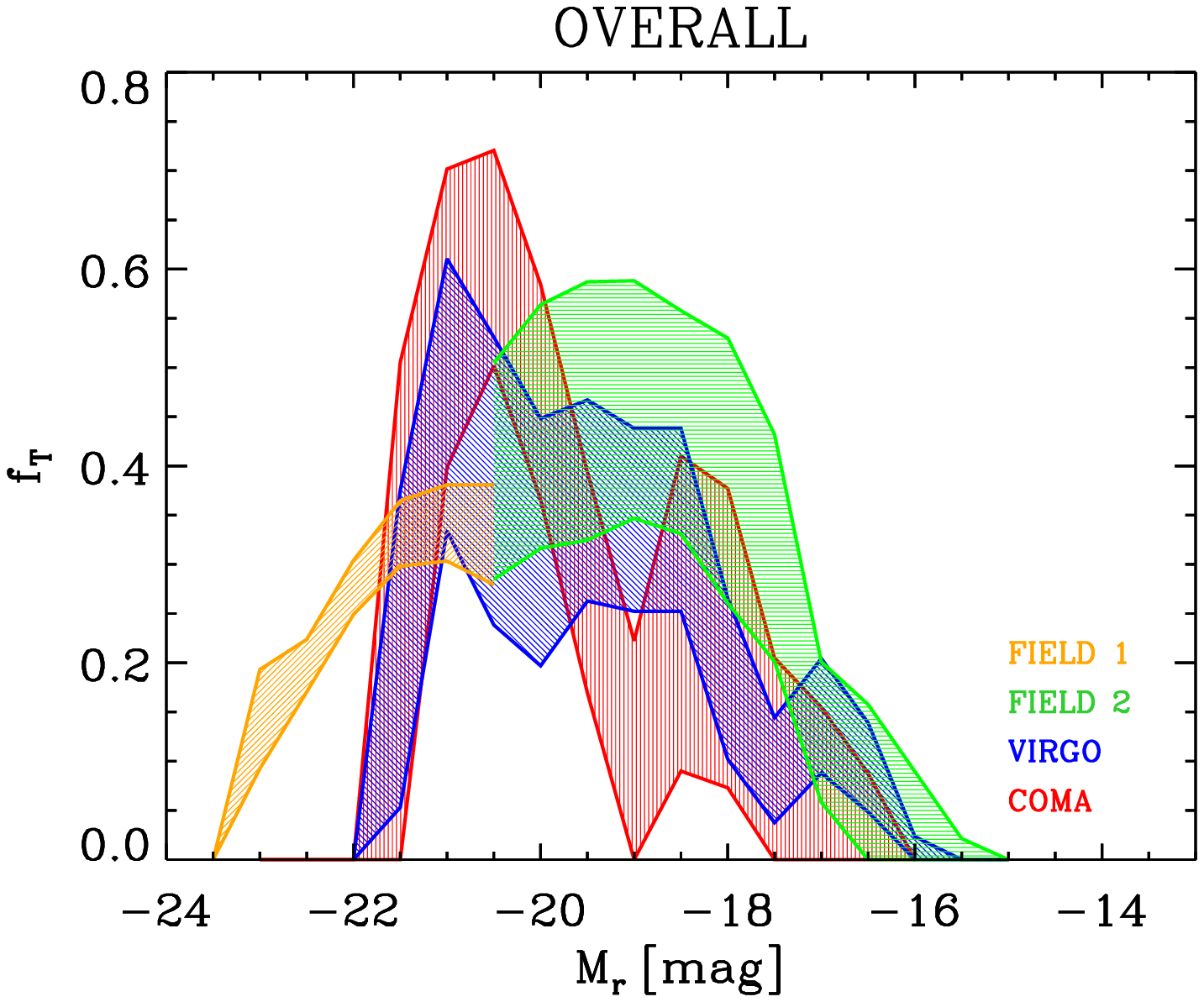}
 \includegraphics[width=0.45\textwidth]{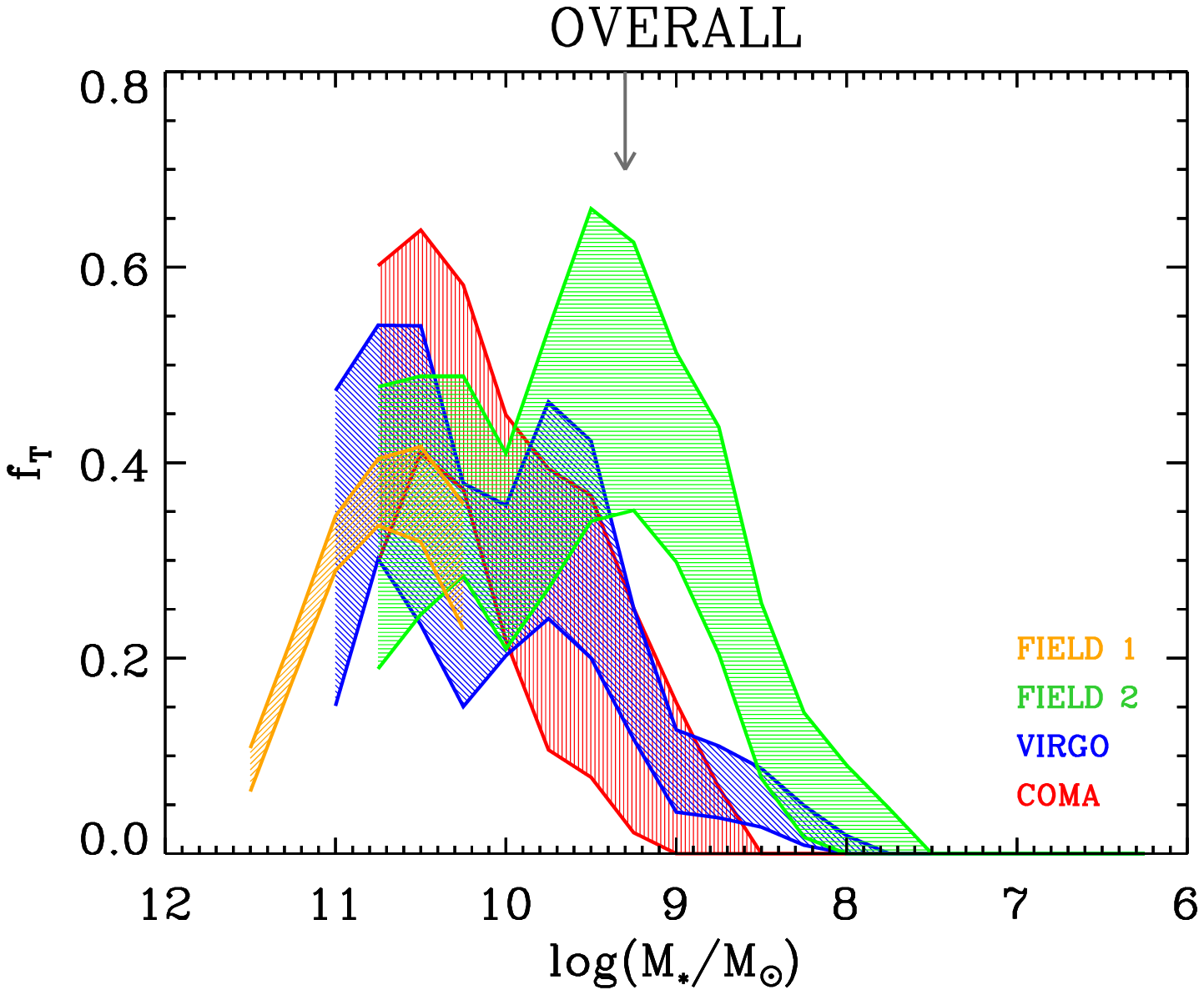}
 \includegraphics[width=0.45\textwidth]{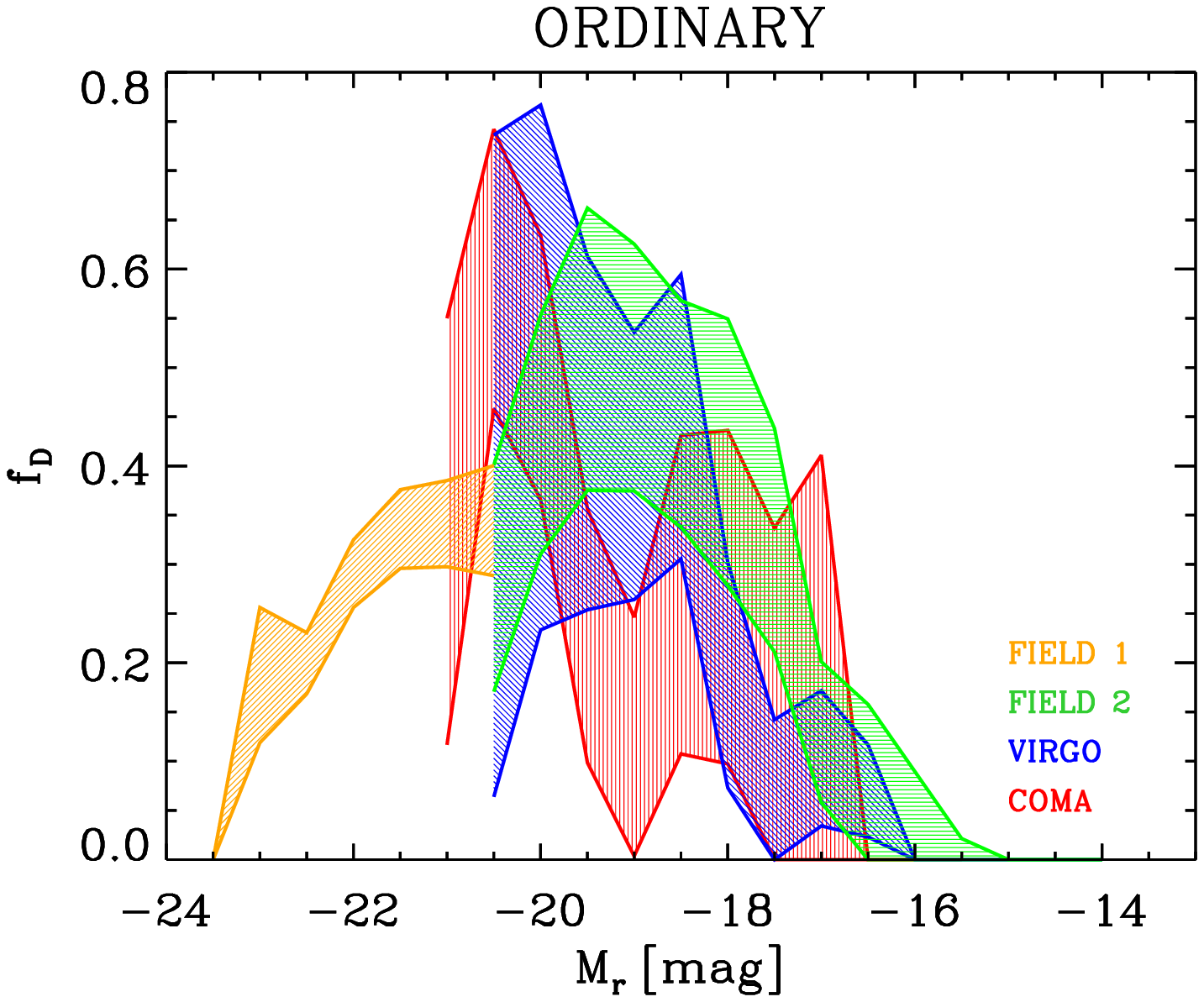}
 \includegraphics[width=0.45\textwidth]{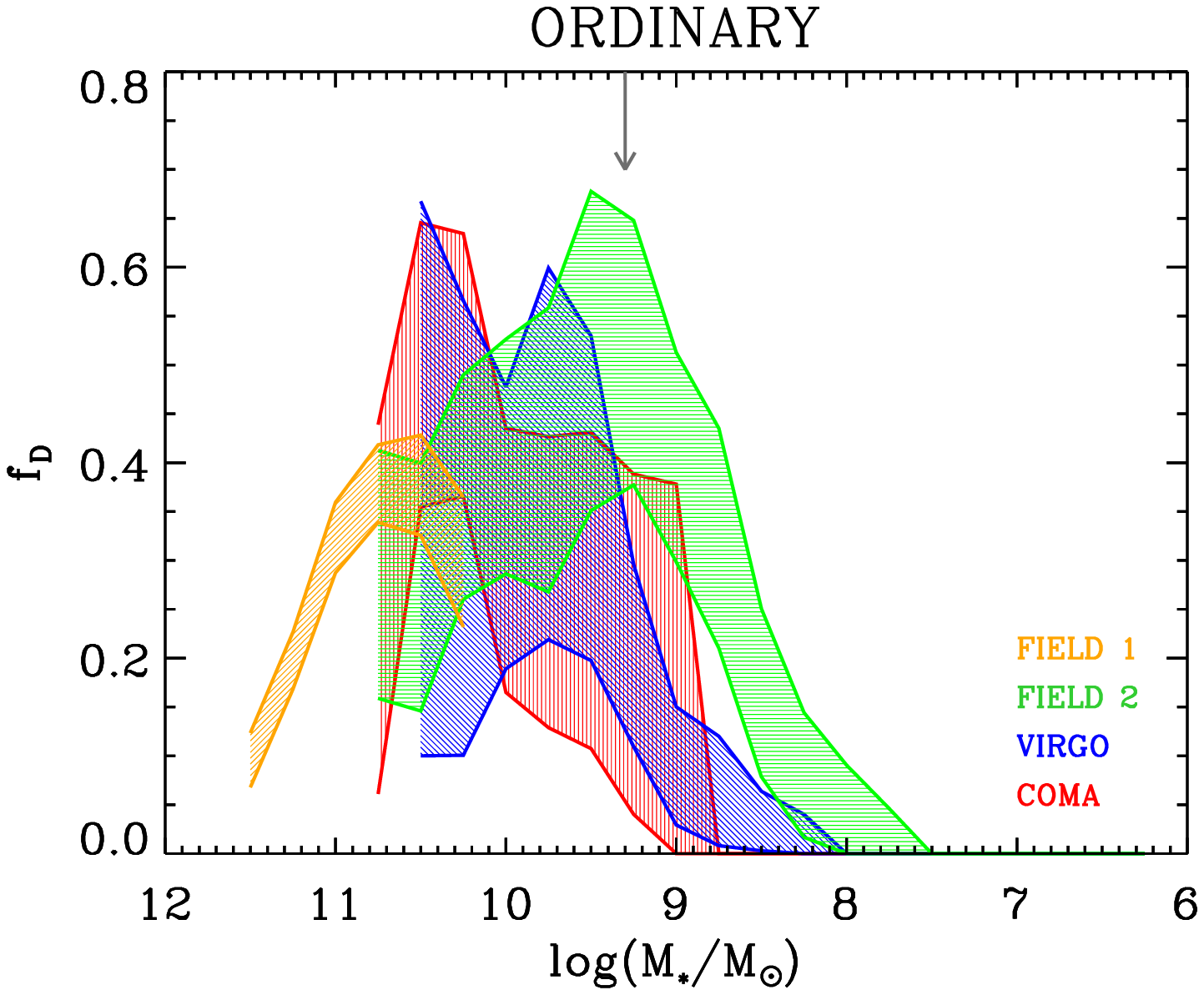}
 \caption{\footnotesize Bar fraction distribution as function of the
   galaxy magnitudes (left panels) and masses (right panels).  The bar
   fraction calculated using all the Hubble types ($f_{\rm T}$) and
   only the disk galaxies ($f_{\rm D}$) are plotted in the upper and
   bottom panels, respectively. The field1, field2, Virgo, and Coma
   samples are showed by the hatched orange, green, blue, and red
   areas, respectively.  The grey arrow indicates the characteristic
   mass below which low-mass galaxies start to be systematically
   thicker \citep{SanchezJanssen2010}.}
 \label{fig:1}
 \end{figure*}

Figure \ref{fig:1} shows $f_{\rm D}$ and $f_{\rm T}$ as function of
the $r$-band absolute magnitude and mass of the galaxies.
The lower and upper boundaries  of the hatched areas correspond to the
bar fraction calculated by  considering only the strong (i.e., secure)
bars and both  the strong and weak (i.e.,  secure and uncertain) bars,
respectively and including their  statistical uncertainties. 
The values of $f_{\rm D}$ and $f_{\rm T}$ for the samples of bright
(field1) and faint field galaxies (field2) are in good agreement.
Therefore, for studying the bar fraction distribution we merged the
two samples into a joint sample of field galaxies (hereafter field
sample).
The fraction of barred galaxies shows a maximum of about $50\%$ at
$M_r\,\simeq\,-20.5$ in clusters, whereas the peak is shifted to
$M_r\,\simeq\,-19$ in the field.

\begin{figure*}[!ht]
 \centering
\includegraphics[width=0.8\textwidth, clip=true]{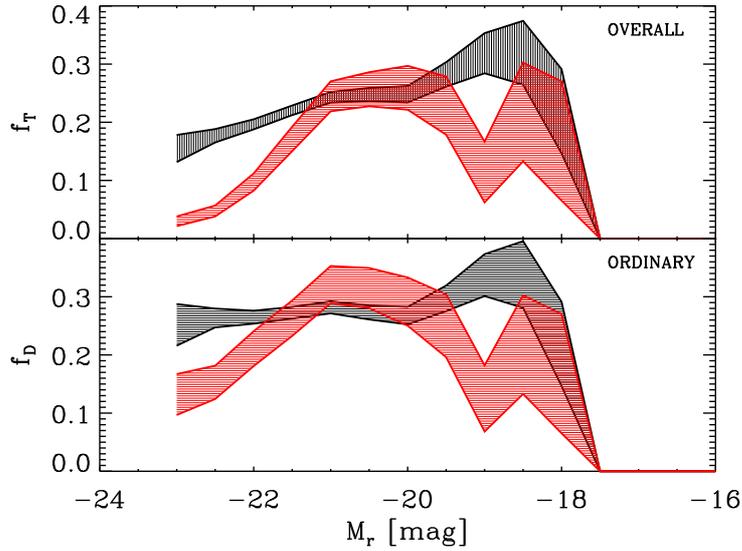}
 \caption{\footnotesize Bar fraction distribution for two subsamples
   of galaxies from \citet{NairAbraham2010} and calculated for all the
   Hubble types (upper panel) and disks only (bottom panel) as a
   function of the galaxy magnitude.  The black and red regions show
   the galaxies in low ($\Sigma_5\,<\,1 $ Mpc$^{-2}$) and high density
   environments ($\Sigma_5\,>\,10$ Mpc$^{-2}$), respectively.}
 \label{fig:2}
 \end{figure*}

For a consistency check we performed the same analysis in the sample
studied by \citet{NairAbraham2010}.  We adopted their morphological
classification and estimate of the local galaxy density to identify
galaxies in the field ($\Sigma_5\,<\,1$ Mpc$^{-2}$) and in compact
groups ($\Sigma_5\,>\,10$ Mpc$^{-2}$). The bar fraction distributions
of these two subsamples probing the very low-density and very
high-density environments, respectively are shown in
Figure~\ref{fig:2}.  The trends we derived from the data by
\citet{NairAbraham2010} are fully in agreement with the findings for
our samples once we take into account that \citet{NairAbraham2010}
impose a more severe definition in identifying barred galaxies and
hence the values of $f_{\rm D}$ and $f_{\rm T}$ we derived for their
samples are systematically lower than those in this work.

\section{Conclusions}

The difference of the bar fraction distributions as a function of
galaxy luminosity (and mass) in the field and Coma cluster is found to
be statistically significant ($>68\%$ confidence level), with the
Virgo cluster being an intermediate case.
Since bars live in disks, these findings allow us to distinguish
between the environmental processes inhibiting bar formation (heating)
or even destroying the host disk from the processes triggering the
disk instabilities which are responsible for bar formation.

We interpret the decrease of $f_{\rm D}$ and $f_{\rm T}$ with
decreasing galaxy luminosity observed for fainter
($M_r\,\gtrsim\,-19$) or, equivalently, less massive galaxies ($ {\cal
  M_{*}}/{\cal M}_{\sun}\,\lesssim\,10^{9}$) as due to the increase of
the disk thickness. Indeed, \citet{SanchezJanssen2010} have recently
found that the minimum of the disk thickness distribution occurs at a
characteristic mass ${\cal M_{*}}/{\cal
  M}_{\sun}\,\simeq\,2\,\times\,10^9$ (corresponding to
$M_r\simeq\,18.5$) below which low-mass galaxies start to be
systematically thicker making it difficult to develop a bar.  We
suggest that the values of $f_{\rm D}$ and $f_{\rm T}$ in the field
are systematically larger than those in Virgo and Coma because the
low-mass galaxy disks in clusters are more easily heated, or
destroyed, by galaxy interactions and can not develop a bar. In the
low luminosity regime nurture and nature are acting on galaxy disks in
cluster and field, respectively.

Since the values of $f_{\rm D}$ and $f_{\rm T}$ are larger for cluster
galaxies with $ -21.5\,\lesssim\,M_r\,\lesssim\,-19.5$ (or $
10^{9.5}\,\lesssim\,{\cal M_{*}}/{\cal M}_{\sun}\,\lesssim\,10^{11}$),
we conclude that brighter disks are strong enough to survive
interactions and form a bar.  In the high luminosity regime nurture
and nature are acting on bar formation in galaxy disks of cluster and
field, respectively.

Our results highlight the importance of studying galaxy samples which
have been carefully selected in luminosity to avoid biases when
dealing with bar statistics. We argue that most of the controversial
results about the relationship between environment and bar fraction
could be explained in terms of the different luminosity ranges covered
by the galaxy samples studied so far.

\bibliographystyle{aa}

\end{document}